\documentclass[fleqn,usenatbib,useAMS]{mnras}
\usepackage{graphicx}	
\usepackage{amsmath}	
\usepackage{amssymb}	
\usepackage{multicol}   
\usepackage{bm}		
\usepackage{pdflscape}	
\usepackage{soul}
\usepackage{appendix}
\usepackage{ulem}
\usepackage{tcolorbox}

\usepackage{multibib}
\newcommand{\beq}{\begin{equation}}
\newcommand{\eeq}{\end{equation}}
\newcommand{\bse}{\begin{subequations}}
\newcommand{\ese}{\end{subequations}}
\newcommand{\bary}{\begin{eqnarray}}
\newcommand{\eary}{\end{eqnarray}}
\newcommand{\bwt}{\begin{widetext}}
\newcommand{\ewt}{\end{widetext}}

\usepackage[T1]{fontenc}
\usepackage{ae,aecompl}

\title{Redshift constrain of BL Lac PKS 1424+240}

\author[Sarira Sahu et al.]
{
Sarira Sahu$^{1}$%
\thanks{Contact e-mail: \href{mailto:sarira@nucleares.unam.mx}{sarira@nucleares.unam.mx}},\ %
D. I. Páez-Sánchez$^{2}$%
\thanks{Contact e-mail: \href{mailto:paez@ciencias.unam.mx}{paez@ciencias.unam.mx}},\ %
B.~Medina-Carrillo$^{3}$%
\thanks{Contact e-mail: \href{mailto:benjamin.medina@cinvestav.mx}{ benjamin.medina@cinvestav.mx}},\ %
R. de J.~Pacheco-Aké$^{3}$%
\thanks{Contact e-mail: \href{mailto:rodrigo.pacheco@cinvestav.mx}{ rodrigo.pacheco@cinvestav.mx}},\ %
\newauthor
{
G.~Sánchez-Colón$^{3}$%
\thanks{Contact e-mail: \href{mailto: gabriel.sanchez@cinvestav.mx}{gabriel.sanchez@cinvestav.mx}},\ %
Subhash Rajpoot$^{4}$%
\thanks{Contact e-mail: \href{mailto: Subhash.Rajpoot@csulb.edu}{Subhash.Rajpoot@csulb.edu}}%
}
\\
$^{1}$Instituto de Ciencias Nucleares, Universidad Nacional Aut\'onoma de M\'exico,\\
Circuito Exterior S/N, C.U., A.P. 70-543, CDMX 04510, México.\\
$^{2}$Facultad de Ciencias, Universidad Nacional Aut\'onoma de M\'exico,   Circuito Exterior, C.U.,  \\ A.P. 70-543, 04510 CDMX, México.\\
$^{3}$Departamento de Física Aplicada, Centro de Investigación y de Estudios Avanzados del IPN, Unidad Mérida.\\
A.P. 73, Cordemex, Mérida, Yucatán 97310, México.\\
$^{4}$Department of Physics and Astronomy, California State University,
1250 Bellflower Boulevard, Long Beach,\\
CA 90840, USA.
}

\date{}

\pubyear{2024}

\begin{document}
\label{firstpage}
\pagerange{\pageref{firstpage}--\pageref{lastpage}}
\maketitle

\begin{abstract}
In the period between 2009 and 2015, several very high-energy (VHE $> 100$ GeV) gamma-ray flaring events from the BL Lac object PKS 1424+240 were observed by the Cerenkov telescopes VERITAS and MAGIC. It had uncertain redshift ($z$) and using spectroscopical measurement, Paiano et al. (2017) found it to be $z=0.604$. Using four different extragalactic background light (EBL) models and the photohadronic model, nine independently observed VHE gamma-ray spectra of PKS 1424+240 are analyzed and a global $\chi^2$ fit is performed on all observations to estimate the best-fit value for the redshift for each EBL model. Confidence levels (CL) intervals for the redshift are also estimated using all the EBL models. This method is tested by comparing our analysis with the observed value. It is shown that the photohadronic scenario provides an excellent description of all the observed spectra. It is found that the EBL model of Domínguez et al. (2011) is the one that provides the most restrictive limits on the redshift of PKS 1424+240, but in our analysis, $z=0.604$ lies within the $3\sigma$ CL interval of the EBL model of Saldana-Lopez et al. (2021).
\end{abstract}

\color{black}
\section{INTRODUCTION}\label{Intro}

Blazars are a subclass of active galactic nuclei (AGN), characterized by their non-thermal spectra, originating from a relativistic jet, that is closely aligned with the observer’s line of sight~\citep{Urry_1995, VERITAS:2010vjk}. Blazars exhibit rapid variability throughout their entire electromagnetic spectrum. Their spectral energy distributions (SEDs) show a distinctive double-peak structure~\citep{Abdo:2009iq}, the first peak arises as synchrotron photons from the propagation of low-energy electrons in the jet magnetic field. However, the second peak can either be from the Compton scattering of high-energy electrons with their self-produced synchrotron photons, a phenomenon known as synchrotron self-Compton (SSC) scattering~\citep{1992ApJ...397L...5M,1993ApJ...416..458D, 1994ApJ...421..153S,Blazejowski:2000ck,Murase_2012,Gao:2012sq} or from the external Compton scattering with the photons from the accretion disk, broad-line regions or the dusty torus\citep{1993ApJ...416..458D,1994ApJ...421..153S,Blazejowski:2000ck}.

PKS 1424+240 was discovered in the 1970s as a radio source~\citep{1974A&AS...18..147F} and was later identified as a blazar by~\cite{1988ApJ...333..666I}. Later, the source was detected in $\gamma$-rays by the \textit{Fermi} Large Area Telescope (LAT;~\cite{2009ApJ...697.1071A}). In the spring of 2009, VERITAS reported the observation of VHE gamma-rays from this source~\citep{2009ATel.2084....1O} and, subsequently, it was confirmed by MAGIC~\citep{2009ATel.2098....1T}. Afterwards, several VHE flaring events from PKS 1424+240 were observed between 2009 and 2015. Nonetheless, the frequency of the synchrotron peak $\nu^{peak}_{sync}$ for PKS 1424+240 was not directly determined. However, it could be constrained from the optical and the X-ray data and was found to be in the range $10^{15}\, \mathrm{Hz} \lesssim \nu^{peak}_{sync}\lesssim 10^{17}$ Hz~\citep{VERITAS:2009lpb}. From the classification of the BL Lac objects, we know that this range of $\nu^{peak}_{sync}$ classifies the object as high-frequency peaked BL Lac (HBL)~\citep{1996MNRAS.279..526P, Abdo:2009iq, Boettcher:2013wxa}. 

The non-thermal emission in a BL Lac object dominates over the stellar emission of its host galaxy, making it difficult to correctly estimate the redshift. As a consequence, ambiguities arise in tracing the cosmic evolution, the nature and the understanding of the intrinsic VHE spectrum of the source. Given the fact that VHE photons are dampened by the extragalactic background light (EBL), information of the redshift is essential to investigating the role of the EBL.

EBL models are constructed with the convergence of different approaches as direct measurements, semi analytical models of galaxy formation, analysis of cosmic star formation history, and galaxy luminosity densities with SEDs. The models used in this work belong to the last category and their main difference is the number of galaxy samples employed, which is correlated with the redshift value range of the data output from the model~\citep{2021MNRAS.507.5144S}.

The VHE gamma-ray flux of a blazar observed on Earth is given by~\citep{Hauser:2001xs}
\beq
F_{\gamma}(E_{\gamma})=F_{int}(E_\gamma)\, e^{-\tau_{\gamma\gamma}(E_{\gamma}, z)}, 
\label{eq:flux}
\eeq
where $E_\gamma$, $F_{\gamma}$, and $F_{int}$ are the observed VHE photon energy, the observed flux, and the intrinsic flux, respectively. Here, the optical depth $\tau_{\gamma \gamma}$ for the process $\gamma\gamma\rightarrow e^+e^-$ depends on $E_\gamma$ and the redshift $z$ of the source. The exponential factor in Eq. (\ref{eq:flux}) accounts for the attenuation in the VHE flux due to $e^+e^-$ production~\citep{1992ApJ...390L..49S,doi:10.1126/science.1227160,Padovani:2017zpf}. Thus, knowledge on $z$ is crucial to estimating the intrinsic flux from the observed flux. Well-known and reliable EBL models by~\cite{Franceschini:2008tp,2010ApJ...712..238F,Dominguez:2010bv,10.1111/j.1365-2966.2012.20841.x,2021MNRAS.507.5144S} are employed by Imagining Atmospheric Cerenkov Telescope (IACT) collaborations to analyze the observed VHE $\gamma$-ray spectra from objects at different redshifts. 

As stated previously, the redshift of HBL PKS 1424+240 is measured spectroscopically with a value $z = 0.604$~\citep{Paiano_2017}. However, alternative methods do not provide a definitive result. By making a ``minimum luminosity assumption" (host galaxy absolute magnitude to be $M = - 21.9$), \cite{1995A&A...303..656S} have derived a lower limit of $z > 0.06$, while~\cite{Sbarufatti_2005} set a limit of $z > 0.67$ by using the average value $M = - 22.9$. \cite{VERITAS:2009lpb} combined the photon index (measured with {\it Fermi} Large Area Telescope) with EBL models, to estimate the redshift $z=0.5 \pm 0.1$, with an upper limit of $z<0.66$ at $95\%$ CL. Comparing the measured and intrinsic VHE spectra due to EBL absorption, an upper limit of 1.19 on the redshift was derived in~\cite{10.1093/pasj/62.4.L23}. Employing a derived empirical law, describing the relation between the upper limits and the true redshifts, the redshift is estimated to be $0.24\pm 0.05$ in~\cite{10.1111/j.1745-3933.2010.00862.x,2011NCimC..34c.241P}. The photometric redshift, obtained through spectral fitting of optical/UV data, has an upper limit of $z<1.11$~\citep{refId0}. \cite{Furniss:2013roa} have given a lower limit, $z\geq 0.6035$, set by the detection of Ly$\beta$ and Ly$\gamma$ lines from intervening hydrogen clouds. By multiwavelength spectral characterization and modeling of the blazar, in Ref.~\cite{MAGIC:2014vpp}, a value $z=0.61\pm 0.10$ with an upper limit of 0.81 at $2\sigma$ is reported. In Ref.~\cite{refId03}, the probability of PKS 1424+240 being a member of a group of galaxies found at $z = 0.6010\pm 0.003$ was found to be $98\%$. The observed correlation between the VHE spectral index of blazars and redshift is interpreted as a result of EBL induced absorption effects and the absence of such correlation led to develop a technique to estimate the redshift of distant blazars whose optical spectrum is featureless. For PKS 1424+240, this technique was employed by \cite{Zahoor:2021frv} and have estimated $z= 0.28 \pm 0.13$ and $z=0.24 \pm 0.11$.

The multi-zone leptonic models are also used to explain the VHE flaring events. However, one needs to almost double the parameters in these models. Models involving hadrons suffer from low efficiency. For example, in the proton synchrotron scenario, the emission of synchrotron photons by protons in the jet environment is suppressed by a factor of $m^{-4}_p$, where $m_p$ is the proton mass. Thus, this process requires ultra-high energy proton flux. Also a strong magnetic field is needed, which is unusual in a blazar jet~\citep{1991A&A...251..723M,MUCKE2001121,2003APh....18..593M}. There are several other alternative models, such as spine-layer structured jet model and the lepton-hadron hybrid models to explain these spectra~\citep{2005A&A...432..401G,2008MNRAS.385L..98T,2019MNRAS.490.2284M,2018A&A...620A.181A}.

Previously, the photohadronic model~\citep{Sahu:2019lwj}, has been successfully used to explain the VHE $\gamma$-ray spectra of various HBLs and extreme HBLs (EHBLs)~\citep{2020ApJ...901..132S,2021ApJ...914..120S,2022MNRAS.515.5235S}. A new classification scheme was proposed~\citep{Sahu_2019} by analyzing the spectral index of several observed VHE $\gamma$-ray spectra from HBLs. In the foregoing, the photohadronic model~\citep{Sahu:2019lwj} along with four well-known EBL models~\cite{2021MNRAS.507.5144S,Franceschini:2008tp,Dominguez:2010bv,10.1111/j.1365-2966.2012.20841.x} are used to analyze nine VHE flaring events of PKS 1424+240 between 2009 and 2015, observed by VERITAS and MAGIC collaborations. By performing a $\chi^2$ global fit to the full set of experimental data points of these VHE spectra, the central values and CL intervals for the redshift of PKS 1424+240 are determined for each EBL model considered.

The plan of this paper is as follows. In Section~\ref{sec3} a brief review of the photohadronic model and its features relevant to the present work is presented. In Section~\ref{sec4} the analysis and results are presented. Finally, Section~\ref{sec5} includes a short summary and discussion.

\section{THE PHOTOHADRONIC MODEL}\label{sec3}

In the photohadronic model, a double jet configuration is assumed during the VHE flaring process~\citep{Sahu:2019lwj,Sahu_2019}. A double jet configuration has been proposed in earlier works~\citep{10.1111/j.1365-2966.2008.13360.x,10.1111/j.1365-2966.2009.16045.x}. In the photohadronic scenario, a compact and constrained narrower jet, with a size $R'_f$, is formed within the wider jet of size $R'_b$, $R'_f < R'_b$ (the quantities in the comoving frame are indicated by primes). In the inner region of the jet, the photon density, $n'_{\gamma,f}$, is significantly greater than the photon density in the outer jet region, $n'_\gamma$ ($n'_{\gamma,f} \gg n'_\gamma$). The photohadronic model is based on the conventional interpretation of the first two peaks in the SED, namely, that the first peak results from the synchrotron radiation emitted by relativistic electrons within the jet environment and the second peak is a consequence of the SSC process. The inner jet moves at a (slightly) higher velocity (with bulk Lorentz factor $\Gamma_{in}$) compared to the outer jet (with bulk Lorentz factor $\Gamma_{ext}$). For simplicity, it is assumed $\Gamma_{in} \simeq \Gamma_{ext}\equiv \Gamma$ and a common Doppler factor $\mathcal{D}$~\citep{Ghisellini:1998it,Krawczynski:2003fq}. For HBLs, we have $\Gamma\simeq {\cal D}$.
 
In the photohadronic scenario, protons are accelerated to very high-energies in the inner jet region, and their differential spectrum follows a power-law form, $dN_p/dE_p\propto E^{-\alpha}_p$~\citep{1993ApJ...416..458D}, where $E_p$ is the proton energy and $\alpha\geq 2$ is the proton spectral index. The specific value of $\alpha$ differs, depending on the type of shock encountered. It can be different for non-relativistic shocks, highly relativistic shocks, and oblique relativistic shocks~\citep{2005PhRvL..94k1102K,2012ApJ...745...63S}. The interaction of these high-energy protons with the SSC background seed photons in the inner jet region results in the production of the $\Delta$-resonance through the process $p+\gamma\rightarrow\Delta^{+}$. The $\Delta$-resonance decays to $\pi^0$ and $\pi^+$ with different probabilities. Although the direct single pion production and the multi-pion production processes contribute, they are less efficient in the energy range under consideration ~\citep{1999PASA...16..160M,2018MNRAS.481..666O}. Such contributions are neglected in the present work. Finally, the neutral pion decays to $\gamma$-rays and the $\pi^+$ to neutrinos~\citep{PhysRevD.85.043012}. In the photohadronic scenario, the $\gamma$-rays generated from $\pi^0$ decay are blue-shifted to VHE $\gamma$-rays and are detected on Earth. The positrons produced from $\pi^+$ decay will emit synchrotron photons.

The observed VHE $\gamma$-ray energy $E_\gamma$ and the seed photon energy $\epsilon_\gamma$ satisfy the condition~\citep{Sahu:2019lwj,Sahu_2019}
\beq
E_{\gamma}\epsilon_{\gamma}\simeq 0.032 \frac{{\cal D}^2}{(1+z)^2}\,\mathrm{GeV}^2.
\label{KinCon}
\eeq
In the above process, the VHE photon carries approximately $10\%$ of the proton energy, $E_p=10E_\gamma$. Given the inaccessible nature of the inner jet region and the lack of direct means to estimate the photon density within it, a scaling relationship between the inner and the outer jet regions is assumed, which can be expressed as~\citep{Sahu:2015tua}
\beq
\frac{n'_{\gamma,f}(\epsilon_{\gamma,1})}{n'_{\gamma,f}(\epsilon_{\gamma,2})} \simeq\frac{n'_{\gamma}(\epsilon_{\gamma,1})}{n'_{\gamma}(\epsilon_{\gamma,2})}.
\label{eq:scalingI}
\eeq
In the above equation, the left-hand side represents the unknown, while the right-hand side is known. This equation is employed to express the unknown photon density in the inner region in terms of the known photon density in the outer region.

The intrinsic $\gamma$-ray flux from the $\pi^0$ decay is deduced to be
\beq
F_{\gamma}(E_{\gamma}) \equiv E^2_{\gamma} \frac{dN(E_\gamma)}{dE_\gamma} 
\propto  E^2_p \frac{dN(E_p)}{dE_p} n'_{\gamma,f}.
\eeq
By taking into account the seed photon density and the proton flux, the intrinsic flux can be expressed as
\beq
F_{int}(E_{\gamma})=F_0 \left ( \frac{E_\gamma}{\rm TeV} \right )^{-\delta+3}.
\label{eq:fluxintrinsic}
\eeq
By putting this in Eq. (\ref{eq:flux}), the observed flux is determined to be
\beq
F_{\gamma}(E_{\gamma}) = F_0 \left ( \frac{E_\gamma}{\rm TeV} \right )^{-\delta+3}
e^{-\tau_{\gamma\gamma}(E_{\gamma},z)},
\label{fluxrelat}
\eeq
where the spectral index $\delta = \alpha + \beta$. The component $\beta$ represents the power spectrum of the seed photons in the low energy  tail region of the SSC spectrum. The normalization factor $F_0$ can be determined from the observed data. The value of $\delta$ is considered as the only free parameter in the photohadronic model~\citep{Sahu_2019}. As can be seen from Eq.(\ref{fluxrelat}), in the photohadronic model the curvature in the spectrum, if any, can be taken care of by the exponential term~\citep{10.1093/mnras/stz943}.

It is important to note that the photohadronic process works well for $E_\gamma\gtrsim 100$~GeV. Below this energy, the leptonic processes such as the electron synchrotron mechanism and the SSC process have the dominant contribution to the multiwavelength SED.

\section{ANALYSIS AND RESULTS}\label{sec4}

In the analysis of the VHE gamma-ray spectra of PKS 1424+240, the EBL corrections are taken care of by the use of four well-known EBL models. For convenience, the EBL models are referred to as Saldana~\citep{2021MNRAS.507.5144S}, Franceschini~\citep{Franceschini:2008tp}, Domínguez~\citep{Dominguez:2010bv}, and Gilmore~\citep{10.1111/j.1365-2966.2012.20841.x}. Also for convenience, each observation is named according to the telescope and the year of observation. The observations by VERITAS in 2009 (from MJD 54881 to MJD 55003) are named as VO09-I for the time averaged spectrum analysis and as VO09-II for the analysis of the VHE $\gamma$-ray spectrum. In Table~\ref{table1} we have listed details about all the observations.

In accordance with the classification scheme given by~\cite{Sahu_2019}, the VHE emission states of a HBL are defined according to the $\delta$ value. The emission state is considered very high when $2.5\leq\delta\leq 2.6$. The high emission state corresponds to $2.6 < \delta < 3.0$ and $\delta=3.0$ implies the low emission state. Since PKS 1424+240 is a HBL, $\delta$ must be constrained to lie in the range $2.5 \leq\delta\leq 3.0$.

The duration of time observed during each observation period by VERITAS and MAGIC are between two to five months. Thus, the observed spectra are the average spectra. It is very important to note that these observation periods are too long for a spectrum to be in a very high emission state or high emission state. Also, the average spectrum of a long observation is always in a low emission state which corresponds to $\delta=3.0$. Thus, for further analysis to all the nine spectra, we fixed the spectral index to $\delta=3.0$.
 
For our analysis and to find the best fits to the observed VHE-spectra for a given EBL model, the EBL model is implemented in the photohadronic model and a global $\chi^2$ fit is performed on all the data points by simultaneously varying the redshift $z$ (common to all observations) and the corresponding normalization constants $F_0$ of each of the nine independent observations. The global fitting procedure is repeated for the other three EBL models too to find the best-fit value of the redshift $z$ of the HBL PKS 1424+240 and the normalization $F_0$ of the photohadronic model. The redshift CL intervals at $1\sigma$, $2\sigma$, and $3\sigma$, are then calculated for each of the four EBL models considered. 

The best fit values for the normalization constants $F_0$, corresponding to each observation and for the four EBL models, are given in Table~\ref{table2}. One can observe that for a given observation, the $F_0$ value is very similar in all the EBL models. This clearly shows that all the EBL models are also similar to each other. Analogously, the best-fit values for the redshift $z$ and their CL intervals at $1\sigma$, $2\sigma$, and $3\sigma$, are calculated which are shown in Table~\ref{table3}. With 42 degrees of freedom (52 experimental data points considered and 10 free parameters), values of minimum $\chi^2$ obtained for the global fit are 61.8, 64.74, 66.47, and 63.51 for the EBL models of Saldana, Franceschini, Domínguez, and Gilmore, respectively. The use of a different EBL model predicts different CL intervals of $z$ for each observed VHE spectrum. The photohadronic model with the EBL model of Domínguez provides the most restrictive intervals on the redshift, $z=0.689$ the central value with $0.650 \leq z \leq 0.724$ at $1\sigma$, $0.622 \leq z \leq 0.748$ at $2\sigma$, and $0.610 \leq z \leq 0.758$, at $3\sigma$. Although, the least restrictive intervals on $z$ are obtained with the EBL model of Saldana, the measured value of $z=0.604$ lies within the $3\sigma$ CL interval of this model.

\begin{figure*}
\centering
\includegraphics[width=7in]{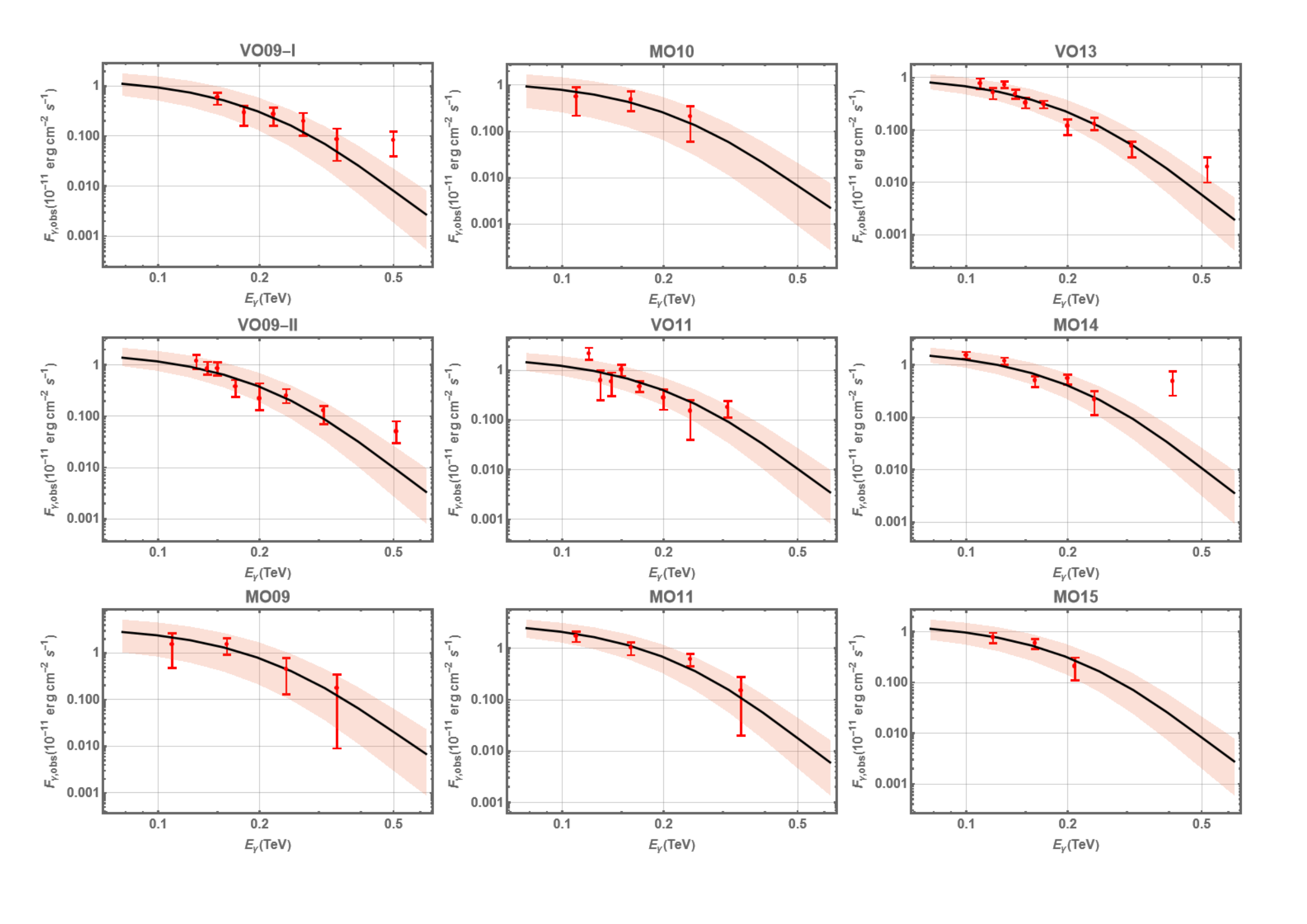}
\caption{Fits to the nine observed VHE $\gamma$-ray spectra of PKS 1424+240 using the photohadronic model with the EBL model of Saldana (solid black curves). All these flaring events are in low emission state corresponding to $\delta=3.0$. Here, $z=0.660$, and the corresponding $F_0$'s are given in Table~\ref{table2}. Plots of the $3\sigma$-CL regions to all the nine observed VHE spectra are also shown.}
\label{fig:Fig1}
\end{figure*}

\begin{figure*}
\centering
\includegraphics[width=5in]{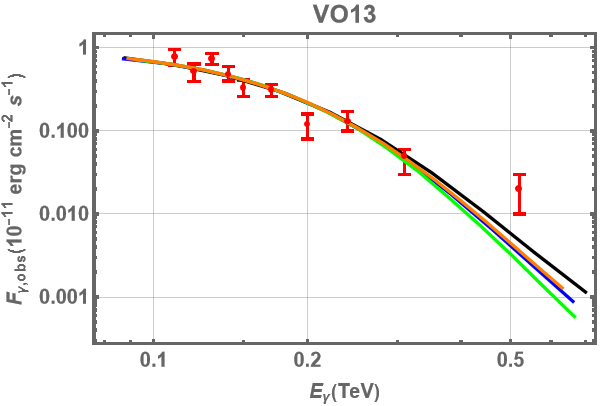}
\caption{Photohadronic model fits to the observed VHE spectrum VO13 of PKS 1424+240 to illustrate the comparison of different EBL models: Saldana (black), Franceschini (blue), Domínguez (green), and Gilmore (orange). Corresponding values of the normalization constants $F_0$ and the redshift $z$ are provided in Tables~\ref{table1} and~\ref{table2}, respectively. As mentioned, for VO13, the VHE flaring is in low emission state, $\delta=3.0$.}
\label{fig:Fig2}
\end{figure*}

Fits to the nine observed VHE spectra of PKS 1424+240 are displayed in Fig.~\ref{fig:Fig1}, where, the best-fit parameters for the EBL model of Saldana (given in Tables~\ref{table2} and \ref{table3}) are used. The fits to the VHE spectra using the Domínguez, Franceschini, and Gilmore EBL models are omitted because, visually, there is no discernible difference from one EBL model to another. For comparison, all the four EBL models are used to best fit the spectrum of VO13 which is shown in Fig.~\ref{fig:Fig2}. All the four EBL models fit very well to the data. However, for $E_{\gamma} \gtrsim 0.3$~TeV we observe a small discrepancy. The Saldana EBL model is implemented in the photohadronic model to determine the butterfly $3\sigma$-CL regions for the nine observed spectra, which are also shown in Fig.~\ref{fig:Fig1}. Plots of CL regions with the other three EBL models considered are omitted. For comparison, the redshift of PKS 1424+240 estimated by different authors and the spectroscopic measurement are summarized in Table~\ref{table4}.

\begin{table*}
\centering
\caption{Information of the PKS 1424+240 observations employed in this work. Name and instrument of the observation is presented in the first and second column, respectively. Start date, end date, period span (in MJD), and duration of the observation are found in the third, fourth, fifth, and sixth columns, respectively. The work in which each observation period data is featured is presented in the seventh column.}
\begin{tabular}{llccccl}
\hline
Observation	& Instrument & Start date & End date & Period (MJD) &	Duration (h) & Reference \\
\hline
VO09-I	&	VERITAS	&	19/02/2009	&	21/06/2009	&	54881 - 55003	&	37.3	&	\cite{VERITAS:2009lpb} \\
VO09-II	&		&		&		&		&		&		\\
MO09	&	MAGIC	&	17/04/2009	&	23/06/2009	&	54938 - 55005	&	12.5	&	\cite{MAGIC:2014vpp} \\
MO10	&	MAGIC	&	14/03/2010	&	19/04/2009	&	55269 - 55305	&	11.6	&	\cite{MAGIC:2014vpp} \\
VO11	&	VERITAS	&	06/02/2011	&	30/05/2011	&	55598 - 55711	&	14	&	\cite{2014ApJ...785L..16A} \\
MO11	&	MAGIC	&   24/04/2011  &	08/05/2011	&   55675 - 55689  &	9.5	&	\cite{MAGIC:2014vpp} \\
VO13	&	VERITAS	&	11/02/2013	&	04/05/2013	&	56334 - 56447	&	67	&	\cite{2014ApJ...785L..16A} \\
MO14	&	MAGIC	&	23/03/2014	&	18/06/2014	&	56739 - 56826	&	28.19	&	\cite{10.1093/mnras/stz943}	\\
MO15	&	MAGIC	&	22/01/2015	&	13/06/2015	&	57044 - 57186	&	20.9	&	\cite{10.1093/mnras/stz943} \\
\hline
\end{tabular}
\label{table1}
\end{table*}

\begin{table*}
\centering
\caption{Estimated values for the normalization constant $F_0$ (in units of $ 10^{-11}\, \mathrm{erg\, cm^{-2}\, s^{-1}}$) of the photohadronic model obtained from the best fits to the VHE spectra of PKS 14+240 for four EBL models are presented in the third column. Refer to the main text for the different observations given in the first column and for the different EBL models mentioned in the second column. The spectral index $\delta=3.0$ (low emission state) is used.}
\begin{tabular}{clc}
\hline
Observation & EBL Model & $F_0$ \\
\hline
VO09-I	&	Saldana	&	1.420 \\
	&	Franceschini	&	1.271 \\
	&	Domínguez	&	1.337	\\
	&	Gilmore&	1.421	\\
VO09-II	&	Saldana	&	1.767	\\
	&	Franceschini	&	1.593	\\
	&	Domínguez	&	1.683	\\
	&	Gilmore	&	1.776	\\
MO09	&	Saldana	&	3.590	\\
	&	Franceschini	&	3.257	\\
	&	Domínguez	&	3.426	\\
	&	Gilmore	&	3.624	\\
MO10	&	Saldana	&	1.191	\\
	&	Franceschini	&	1.074	\\
	&	Domínguez	&	1.126	\\
	&	Gilmore	&	1.194	\\
VO11	&	Saldana	&	1.840	\\
	&	Franceschini	&	1.649	\\
	&	Domínguez	&	1.736	\\
	&	Gilmore	&	1.840	\\
MO11	&	Saldana	&	3.162	\\
	&	Franceschini	&	2.855	\\
	&	Domínguez	&	2.996	\\
	&	Gilmore	&	3.175	\\
VO13	&	Saldana	&	1.034	\\
	&	Franceschini	&	0.931	\\
	&	Domínguez	&	0.981	\\
	&	Gilmore	&	1.037	\\
MO14	&	Saldana	&	1.907	\\
	&	Franceschini	&	1.709	\\
	&	Domínguez	&	1.797	\\
	&	Gilmore	&	1.913	\\
MO15	&	Saldana	&	1.462	\\
	&	Franceschini	&	1.312	\\
	&	Domínguez	&	1.381	\\
	&	Gilmore	&	1.460	\\
\hline
\end{tabular}
\label{table2}
\end{table*}

\begin{table*}
\centering
\caption{Estimated values for the redshift in the photohadronic model obtained from the global $\chi^2$ fit to the VHE spectra of the HBL PKS 1424+240. Refer to the main text for the different EBL models mentioned in the first column. The redshift $z$ is given in the second column. The redshift CL intervals at $1\sigma$, $2\sigma$, and $3\sigma$ are shown in the third, the fourth, and the fifth columns for each EBL model employed, respectively.}
\begin{tabular}{lcccc}
\hline
EBL Model & Redshift & \multicolumn{3}{c}{Redshift CL intervals} \\
 & $z$ & $1\sigma$ & $2\sigma$ & $3\sigma$ \\
\hline
Saldana	& 0.660	& (0.626, 0.724) & (0.607, 0.756) & (0.597, 0.770) \\
Franceschini & 0.679 & (0.639, 0.722) & (0.616, 0.749) & (0.605, 0.762) \\
Domínguez & 0.689 & (0.650, 0.724) & (0.622, 0.748) & (0.610, 0.758) \\
Gilmore	& 0.727 & (0.685, 0.770) & (0.661, 0.798) & (0.650, 0.812) \\
\hline
\end{tabular}
\label{table3}
\end{table*}

\begin{table*}
\centering
\caption{Summary of the spectroscopic measurement and estimations by different authors for the redshift value $z$ of PKS 1424+240. The redshift measurement and estimated interval limits, or central values with uncertainties, are given in the second column, the third column indicates the method used to determine the redshift. Corresponding references are given in the fourth column.
}
\begin{tabular}{clll}
\hline
n & Redshift & Method & Reference \\
\hline
1	&	$z = 0.604$	& Spectroscopic measurement &	\mbox{\cite{Paiano_2017}} \\
2	&	0.06 < $z$	& Minimum luminosity assumption &	\mbox{\cite{1995A&A...303..656S}} \\
3	&	0.67 < $z$	& Average luminosity value &	\mbox{\cite{Sbarufatti_2005}} \\
4	&	$z$ = 0.5 ± 0.1	& Photon index -- EBL models combination &	\mbox{\cite{VERITAS:2009lpb}}	\\
5	&	$z$ < 0.66 at 95\% CL & Photon index -- EBL models combination &	\mbox{\cite{VERITAS:2009lpb}}	\\
6	&	$z$ < 1.19	& Measured -- intrinsic VHE spectra comparison &	\mbox{\cite{10.1093/pasj/62.4.L23}}	\\
7	&	$z$ = 0.24 ± 0.05	& Upper limits -- true redshifts relationship &	\mbox{\cite{10.1111/j.1745-3933.2010.00862.x,2011NCimC..34c.241P}}	\\
8	&	$z$ < 1.11	& Spectral fitting of optical/UV data &	\mbox{\cite{refId0}}	\\
9	&	0.6035 < $z$	& Ly$\beta$ and Ly$\gamma$ lines detection &	\mbox{\cite{Furniss:2013roa}} \\
10	&	$z$ = 0.61 ± 0.10	& Multiwavelength spectral characterization &	\mbox{\cite{MAGIC:2014vpp}}	\\
11	&	$z$ < 0.81 at $2\sigma$	& Multiwavelength spectral characterization &	\mbox{\cite{MAGIC:2014vpp}}	\\
12	&	$z$ = 0.6010 ± 0.003 & Member of a group of galaxies probability &	\mbox{\cite{refId03}} \\
13	&	$z$ = 0.28 ± 0.13	&VHE spectral index -- redshift correlation &	\mbox{\cite{Zahoor:2021frv}} \\
14	&	$z$ = 0.24 ± 0.11	& VHE spectral index -- redshift correlation &	\mbox{\cite{Zahoor:2021frv}} \\
15	&	0.597 < $z$ < 0.770 at 95\% CL	&  Global fit to independent observations &	Present work (EBL Saldana) \\
\hline
\end{tabular}
\label{table4}
\end{table*}

\section{SUMMARY AND DISCUSSION}\label{sec5}

In this work, the photohadronic model is used in conjunction with the EBL models of Saldana, Franceschini, Domínguez, and Gilmore, to impose stringent constraints on the redshift of PKS 1424+240. This is done by analyzing nine different VHE spectra, independently observed by VERITAS and MAGIC telescopes in the period between 2009 and 2015. Since the photohadronic model~\citep{Sahu:2019lwj,Sahu_2019} has demonstrated its effectiveness in constraining the redshift of various HBLs with unknown $z$ values~\citep{2023MNRAS.522.5840S}, it is employed again to analyze the nine VHE spectra of PKS 1424+240 to constrain its redshift.

The observation period for each spectrum varies between two to five months, too long for a spectrum to be in a very high emission state or in a high emission state. It is observed that all the nine VHE emissions of the HBL PKS 1424+240 were in low emission states ($\delta=3.0$). With this information, for the analysis, a global $\chi^2$ fit is performed by simultaneously varying the redshift $z$ of the object and the normalization constants $F_0$ of each one of the nine independent observations, to find the best fit to the observed VHE spectra for a given EBL model. This procedure is repeated for all the other three EBL models to find the corresponding best fit values of $z$ and the parameters $F_0$. Using the best fit values of the redshift $z$ and the $F_0$'s, the redshift CL intervals at $1\sigma$, $2\sigma$, and $3\sigma$, for each one of the four EBL models considered are calculated.

Our analysis shows that the EBL model of ~\cite{Dominguez:2010bv} provides the most restrictive limits on the redshift and the least restrictive intervals on $z$ are obtained with the EBL model of~\cite{2021MNRAS.507.5144S}. However, the measured value of $z=0.604$~\citep{Paiano_2017} lies within the $3\sigma$ CL interval of the EBL model of~\cite{2021MNRAS.507.5144S} only.

\section*{Acknowledgements}
We thank the reviewer for her/his constructive remarks which helped us to improved the manuscript substantially. The work of S.S. is partially supported by DGAPA-UNAM (México) Project No. IN103522. B. M-C, R. de J. P-A, and G. S-C, would like to thank CONAHCYT (México) for partial support. Partial support from CSU-Long Beach is gratefully acknowledged. 


\section*{Data Availability}
No new data were generated or analysed in support of this research.


\bibliographystyle{mnras}
\bibliography{ref}


\bsp	
\label{lastpage}
\end{document}